\documentclass{elsart4-1}

\usepackage{amssymb}
\usepackage{graphicx}
\usepackage[english,francais]{babel}

\newcommand{\lphi}{\ell_{\phi}}
\newcommand{\ns}{n_{\textup{s}}}
\newcommand{\Vg}{V_{\textup{g}}}
\newcommand{\Te}{T_{\textup{e}}}
\newcommand{\SD}{S_{\textup{D}}}

\newtheorem{e-proposition}[theorem]{Proposition}

\newtheorem{e-definition}[theorem]{Definition\rm}

\renewcommand{\theequation}{\arabic{equation}}
\def\og{\leavevmode\raise.3ex\hbox{$\scriptscriptstyle\langle\!\langle$~}}
\def\fg{\leavevmode\raise.3ex\hbox{~$\!\scriptscriptstyle\,\rangle\!\rangle$}}

\begin{document}

\centerline{Physics or Astrophysics/Header}
\begin{frontmatter}


\selectlanguage{english}
\title{Thermoelectric and electrical transport in mesoscopic two-dimensional electron gases}


\selectlanguage{english}
\author[authorlabel1]{Vijay Narayan},
\ead{vn237@cam.ac.uk}
\author[authorlabel2]{Michael Pepper},
\ead{michael.pepper@ucl.ac.uk}
\author[authorlabel1]{David A. Ritchie}
\ead{dar11@cam.ac.uk}

\address[authorlabel1]{Department of Physics, University of Cambridge, J J Thomson Avenuve, Cambridge,CB3 0HE, UK }
\address[authorlabel2]{Department of Electronic and Electrical Engineering, University College London,
Torrington Place, London WC1E 7JE, United Kingdom}


\medskip
\begin{center}
{\small Received *****; accepted after revision +++++}
\end{center}

\begin{abstract}
We review some of our recent experimental studies on low-carrier concentration, mesoscopic two-dimensional electron gases (m2DEGs). The m2DEGs show a range of striking characteristics including a complete avoidance of the strongly localised regime even when the electrical resistivity $\rho >> h/e^2$, giant thermoelectric response, and an apparent decoupling of charge and thermoelectric transport. We analyse the results and demonstrate that these observations can be explained based on the assumption that the charge carriers retain phase coherence over the m2DEG dimensions. Intriguingly, this would imply phase coherence on lengthscales of up to 10~$\mu$m and temperature $T$ up to 10~K which is significantly greater than conventionally expected in GaAs-based 2DEGs. Such unprecedentedly large phase coherence lengths open up several possibilities in quantum information and computation schemes.

{\it To cite this article: V. Narayan, M. Pepper, D. A. Ritchie, C. R. Physique 6 (2005).}

\vskip 0.5\baselineskip

\selectlanguage{francais}
\noindent{\bf R\'esum\'e}
\vskip 0.5\baselineskip
\noindent
{\bf Here is the French title. }
Your r\'esum\'e in French here.
{\it Pour citer cet article~: A. Name1, A. Name2, C. R.
Physique 6 (2005).}

\keyword{Keyword1; Keyword2; Keyword3 } \vskip 0.5\baselineskip
\noindent{\small{\it Mots-cl\'es~:} Mot-cl\'e1~; Mot-cl\'e2~;
Mot-cl\'e3}}
\end{abstract}
\end{frontmatter}

\selectlanguage{francais}
\section*{Version fran\c{c}aise abr\'eg\'ee}

\selectlanguage{english}
\section{Introduction}
\label{Introduction}
The two-dimensional electron gas (2DEG) has served as one of the most versatile arenas to realise and study mesoscopic systems for over three decades now. Mesoscopic systems are systems which are comparable in spatial extent to the electronic `phase-coherence' length $\lphi$, i.e., the length over which the phase of the electron is completely randomised through inelastic processes. These are ideal venues to study fundamental quantum effects such as localisation, but equally, are becoming increasingly important towards the next generation of quantum-based communications and information schemes which rely on the quantum nature of charge carriers. In this context, high-quality GaAs-based 2DEGs in conjunction with state-of-the-art lithographical techniques offer immense scope to create and manipulate mesoscopic structures such as one-dimensional quantum wires~\cite{Wharam_etal_JPhysC1988,vanWees_etal_PRL1988} and zero-dimensional quantum dots~\cite{Smith_etal_JPhysC1988,vanWees_etal_PRL1989}. However, while the electrical and thermoelectric characteristics of quantum wires and quantum dots have been intensely studied, the more basic `mesoscopic 2DEG', i.e., a 2D phase-coherent system, has received much less attention. Typically, 2DEGs have been studied in the macroscopic limit where the system size $L >> \lphi$, giving rise to the so-called `2D metal-to-insulator' (MIT) transition~\cite{Simmons_etal_PRL2000,Hamilton_etal_PRL2000,Abrahams_etal_RMP2001} and integer and fractional quantum Hall effects~\cite{IQHE,FQHE}. In this Review we focus on some of our recent works on 2DEGs of spatial extent $\lesssim~10~\mu$m where both the electrical and thermoelectrical transport contain striking signatures of phase coherent behaviour \textit{even at temperature $T$ = 10~K}. These include (1) a systematic dependence of $\sigma$ on $L$~\cite{BackesPRB2015}, (2) an apparent breakdown of the canonical Mott relation between the electrical conductivity $\sigma$ and Seebeck coefficient (or thermopower) $S$~\cite{NarayanNJP2014}, and (3) a giant enhancement in the magnitude of $S$~\cite{NarayanPRB2012}. In Section~\ref{Discussion} we explain how these striking observations can be understood as arising due to phase coherent transport. Remarkably and very surprisingly, this would imply that $\lphi$ is two orders of magnitude larger than theoretically expected~\cite{Heinzel}, a point we discuss further in Section~\ref{Discussion}.
The remainder of this Review is structured as follows: In Section~\ref{Experimental system} we describe the experimental system and the thermoelectric measurement technique, Section~\ref{Results} constitutes the main body of this article where we review the experimental results, and and finally in Section~\ref{Discussion} we conclude with a discussion and mention of some of the outstanding issues.

\section{Experimental system}
\label{Experimental system}

\begin{figure}
	\centering
		\includegraphics[width = 15cm]{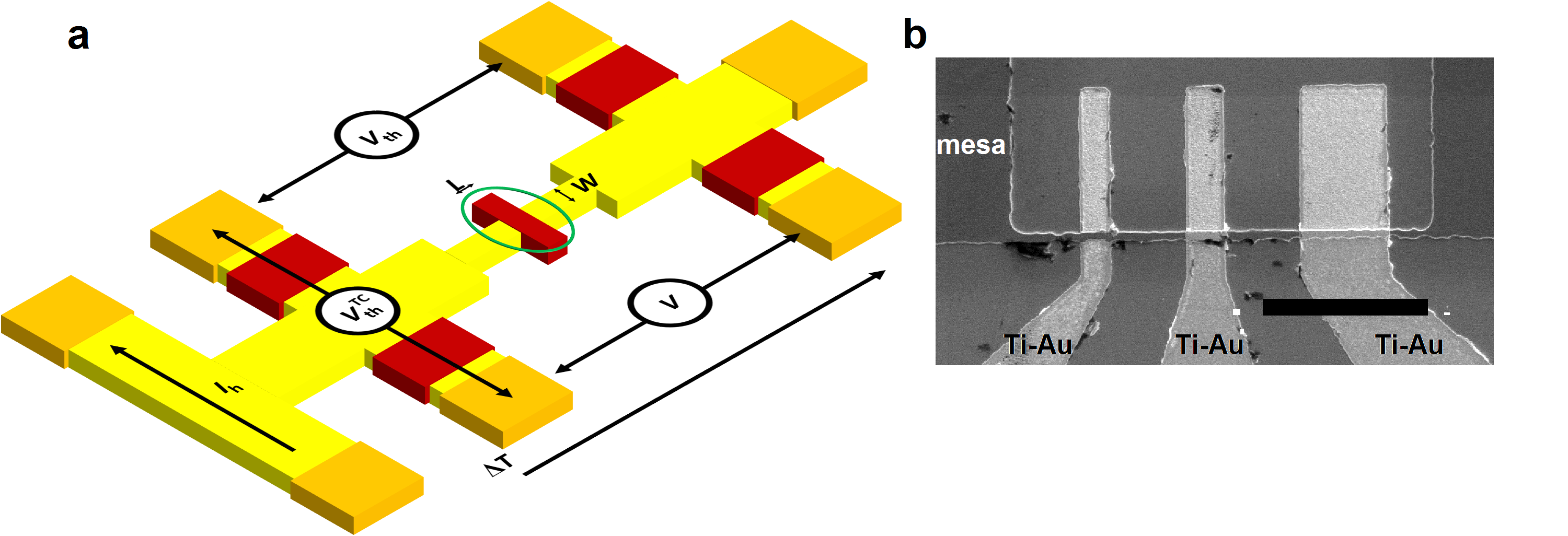}
		\caption{(a) schematic representation of devices. The yellow regions show the conducting mesa with the ohmic contacts depicted in gold. Top-gate electrodes are shown in red. The m2DEG of dimensions $L \times W$ is highlighted in a green ellipse. A heating current $I_{\mbox{h}}$ establishes a temperature gradient $\Delta T$ along the direction shown. The local electron temperature is measured at the locations marked $T_{\mbox{e1}}$ and $T_{\mbox{e2}}$ using lithographically-defined thermocouples extending outwards from the specified locations. The thermovoltage $V_{\mbox{th}}$ in response to $I_{\mbox{h}}$ is measured across the m2DEG in parallel to $V$ which is for the measurement of $\rho$. (b) An SEM of three m2DEGs with varying $L$ defined in a channel with $W = 3~\mu$m. The scale bar shows 20~$\mu$m.}
	\label{F1}
\end{figure}

The 2DEGs described here are realised in $\delta$-doped GaAs heterostructures in which the 2DEG forms $\approx$~300~nm below the wafer surface and the dopants are situated 40~nm above the 2DEG. The as-grown mobility of the wafers was measured at 4~K to be 212~m$^2$/Vs at a carrier density $\ns = 2.2\times10^{12}$~m$^{-2}$. The mesoscopic 2DEGs (m2DEGs) shown in Fig.~\ref{F1} were lithographically patterned as follows: a wet etch was used to define a long channel with width $W$, after which Au-Ge-Ni ohmic contacts were deposited and annealed to make contact to the buried 2DEG. Finally, Ti-Au electrodes were deposited over the channel to create a gate-defined mesoscopic region of size $L \times W$, where $L$ is the length along the transport direction. A gate voltage $\Vg$ applied to the electrodes enabled the tuning of $\ns$ in the m2DEG. The mapping from $\Vg$ to $\ns$ was obtained by examining reflections of edge-states when the device was in the quantum Hall state. Due to the small $L$ ($\lesssim 10~\mu$m) the electrical and thermoelectrical properties were measured in a quasi-four-terminal setup, i.e., as shown in Fig.~\ref{F1}, in addition to the m2DEG there are sections of ungated 2DEG between the voltage leads. However, these additional sections are largely inconsequential since their resistance $R$ is at least an order of magnitude lower than that of the m2DEG in the parameter range of interest. A detailed discussion of this can be found in Ref.~\cite{BackesPRB2015}.

Figure~\ref{F1}a schematically shows the measurement setup: both electrical and thermoelectric properties were measured using AC methods. For the former, an excitation current $I_{\mbox{ex}}$~=~100~pA at a frequency $f \sim 7$~Hz was used. In order to measure the thermoelectric response of the m2DEGs, an AC heating current $I_{\mbox{h}}$ at frequency $f_{\mbox{h}} = 11$~Hz was used to establish a temperature gradient along $L$ which, being $\propto I_{\mbox{h}}^2$, alternates at $2f_{\mbox{h}}$. Thus the thermovoltage $V_{\mbox{th}}$ was measured by locking-in to the second harmonic of $I_{\mbox{h}}$. The measurement of the resultant temperature difference $\Delta T$ across the m2DEG is described in detail in Refs.~\cite{NarayanPRB2012} and \cite{BillialdAPL2015}. Briefly, this was done closely following the scheme initially demonstrated in Ref.~\cite{Syme_etal_JPhys1989,ChickeringPRL2009} where lithographically defined thermocouples were used to measure the local electron temperature. As shown in Fig.~\ref{F1}a each thermocouple consisted of two long and gated 2DEG arms extending outwards from a location along the heated 2DEG and terminating in a ohmic contact each. The heated 2DEG area served as the `hot junction' for a thermocouple and applying a differential voltage bias on the arms resulted in a measurable thermovoltage $V_{\textup{th}}^{TC}$ between the ohmic contacts. $V_{\textup{th}}^{TC}$ was related to the local electron temperature $\Te$ as:

\begin{equation}
\label{LET}
\Te^2 = V_{\textup{th}}^{TC} \frac{6e\hbar^2}{\pi k_B^2 m (1 + \alpha)}\left(\frac{1}{n_1} - \frac{1}{n_2} \right)^{-1} + T_L^2
\end{equation}

Here $m$ is the effective electron mass in GaAs (= 0.067 $m_0$, where $m_0$ is the bare electron mass), $\alpha$ is given by $(\ns/\tau)(d\tau/d\ns)$ where $\tau$ is the elastic scattering time, $n_1$ and $n_2$ are the 2DEG densities beneath the two gated sections of the thermocouple respectively, and $T_L$ is the lattice temperature obtained from a ruthenium oxide thermometer attached to the cold finger of the cryostat. Equation~\ref{LET} is arrived at by invoking the Mott formula~\cite{MottDavis} which states that the diffusion thermopower or Seebeck coefficient $\SD$ of a 2D conductor and its electrical conductivity $\sigma$ are related as:

\begin{equation}
\label{MottGeneral}
\SD = \frac{\pi^2 k_B^2 T}{3 q}\left ( \frac{d \ln \sigma}{d E} \right)_{E = \mu}
\end{equation}

where $q$ is the charge per unit carrier, $E$ is the energy and $\mu$ the chemical potential of the 2D conductor. In the specific case when $\sigma = n_s e^2 m/\tau$, where $-e$ is the electronic charge, Equation~\ref{MottGeneral} reduces to:

\begin{equation}
\label{Mott}
\SD = \frac{\pi k_B^2 T m}{3 e \hbar^2} \frac{1 + \alpha}{\ns}
\end{equation}

Equation~\ref{Mott} provides an analytical expression for $S$ in a non-interacting, classical 2DEG, i.e., in which quantum and interaction effects are imperceptible, and we will use it as a benchmark to gauge the contribution due to elecron-electron correlations and/or localisation.

\section{Experimental Results}
\label{Results}

\begin{figure}
	\centering
		\includegraphics[width = 15cm]{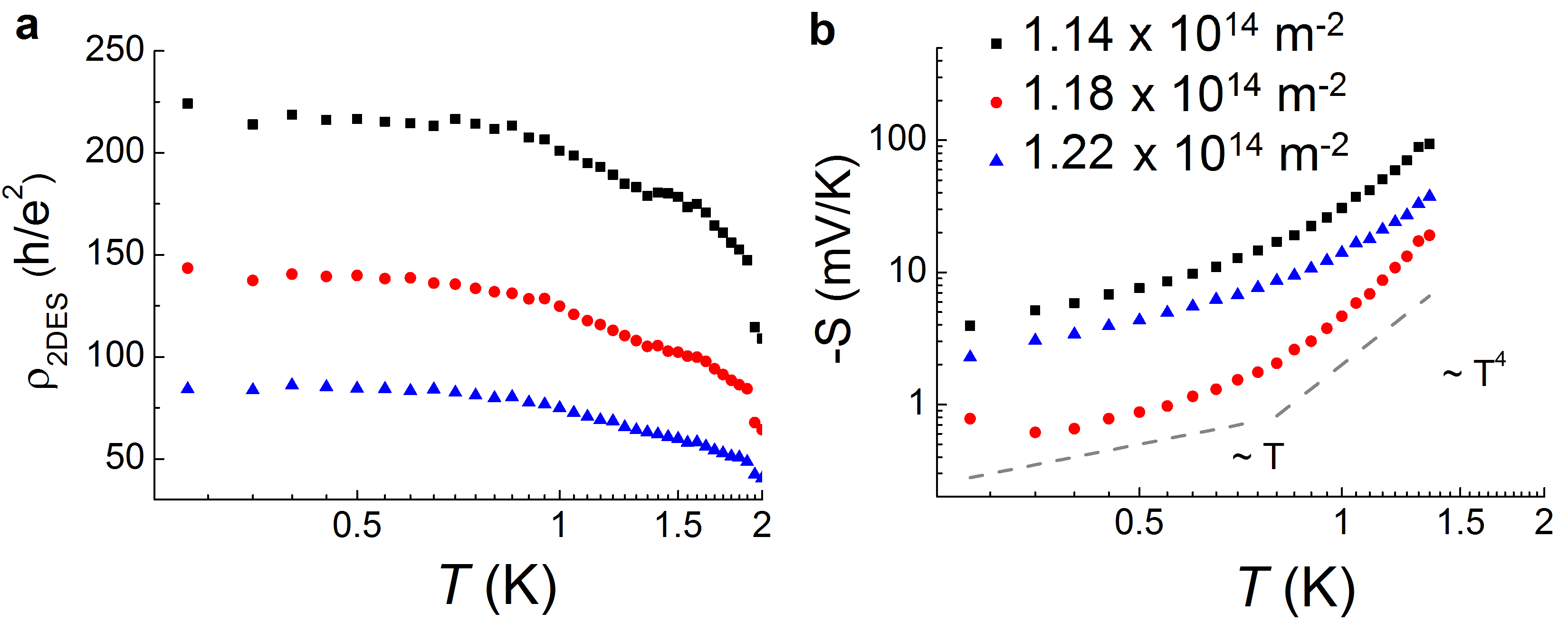}
		\caption{(a) Even when nominally in the strongly localised regime the m2DEGs do not show activated-type behaviour characteristic of hopping transport. Instead, $\rho$ is almost completely $T$-independent below $\approx 1$~K. (b) In the same regime, $S$ shows manifestly metal-like characteristics, albeit being orders of magnitude larger than conventionally expected (see Eq.~\ref{Mott}). Figure adapted from Ref.~\cite{NarayanPRB2012}.}
	\label{F2}
\end{figure}

\begin{figure}[b]
	\centering
		\includegraphics[width = 15cm]{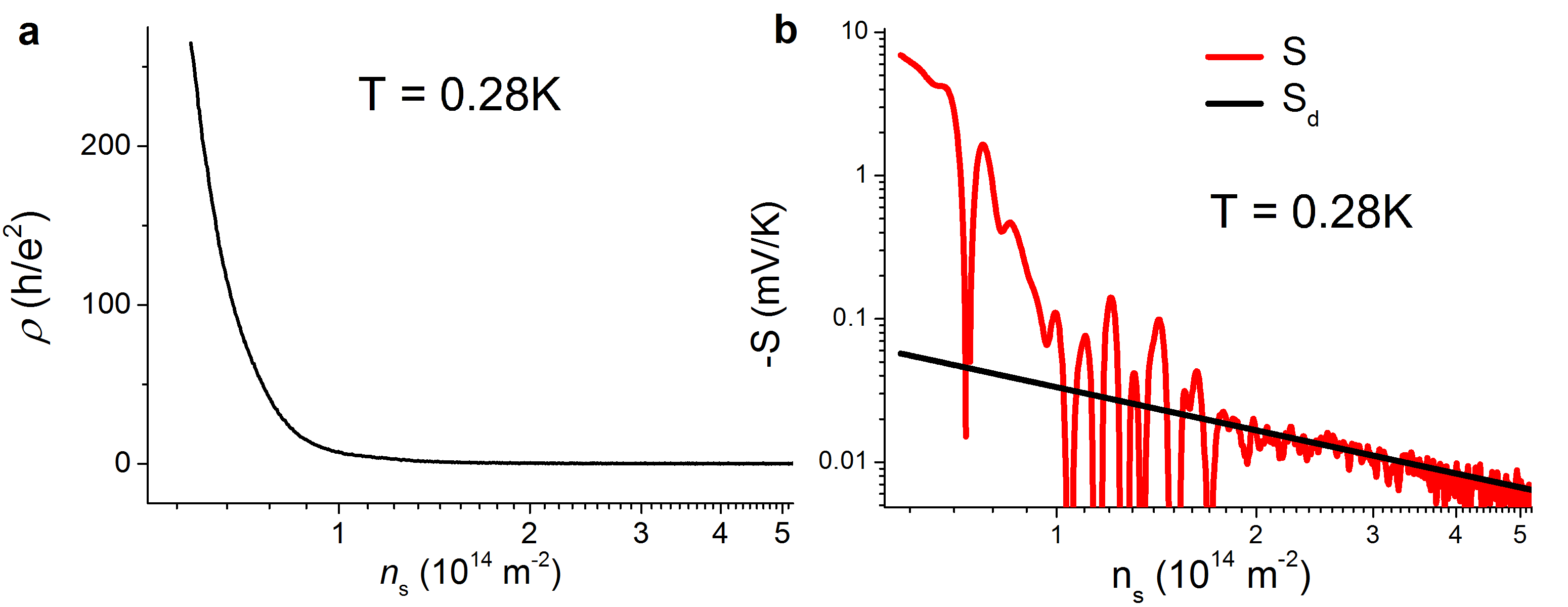}
		\caption{The $\ns$-dependences of $\rho$ and $S$ are apparently in striking disagreement with the expectations of Eq.~\ref{MottGeneral}. While $\rho$ is essentially monotonic in $\ns$, $S$ shows sharp oscillations and even sign-reversals. Figure adapted from Ref.~\cite{NarayanJLTP2013}.}
	\label{F3}
\end{figure}

In this section we will describe the electrical and thermoelectrical transport properties of m2DEGs. However, in ordert to appreciate the striking nature of experimental findings, we begin with a brief note on the behaviour of macroscopic 2DEGs. High-$\ns$ 2DEGs in both Si-based MOSFETs and GaAS-based heterostructures show many outwardly metallic transport characteristics, most notable of which are $\mbox{d}\rho/\mbox{d}T \geq 0$~\cite{Simmons_etal_PRL2000,Hamilton_etal_PRL2000,Abrahams_etal_RMP2001} and a linear dependence of $S$ on $T$~\cite{Fletcher_etal_SST2001}. While the existence of a 2D metallic phase would be in direct contradiction to the expectations of the scaling hypothesis of localisation~\cite{AbrahamsetalPRL1979,LeeRamakrishnanRMP1985}, most if not all of the features of the the `metallic' phase can be understood within the framework of weak localisation (WL) as arising due to finite (but small) $\lphi$. This in turn would imply that the true ground state of the 2DEG is electrically insulating, consistent with the expectations of the scaling hypothesis. As $\ns$ is lowered, however, macroscopic 2DEGs are universally observed to crossover to the `Anderson localised' or `strongly localised' regime at $\rho \approx h/e^2$ in which electron transport is via phonon-assisted `hops'. The $\ns$ at which this so-called `2D MIT' occurs is non-universal and in the range of $10^{14}$~m$^{-2}$, below which $\rho \sim \exp{(\Delta/k_{\mbox{B}}T)}$, with $\Delta$ being the hopping energy. The 2D MIT has been the subject of much debate, the central point of contention being whether the experimental data is indicative of a true $\ns$-driven quantum phase transition, or simply a disorder-driven crossover~\cite{SDSarma_etal_PRL2005}. Crucially, however, regardless of the underlying mechanism, macroscopic 2DEGs are \textit{always} observed to be in the strongly localised regime when $\rho \gtrsim h/e^2$, with both $\rho$ and $S$ diverging as $T \rightarrow 0$~K. This is perhaps, the most striking distinction between macroscopic and mesoscopic 2DEGs.

Figure~\ref{F2} shows the low-$T$-dependence of $\rho$ and $S$ of m2DEGs in the low-$\ns$  regime where $\rho >> h/e^2$. In sharp contrast to macroscopic samples, and as was consistently reported in a number of works on m2DEGs (Refs.~\cite{BaenningerPRL2008,RKoushik_etal_PRB2011,NarayanPRB2012,NarayanJLTP2013,NarayanNJP2014,BackesPRB2015,BackesJPCM2016})	, the activated growth of $\rho(T)$ shows a striking slowing down below $\approx 1$~K. More precisely, $\mbox{d}\rho/\mbox{d}T$ is found to be $\geq 0$ for $T \lesssim 1$~K, with $\rho$ appearing to saturate to a large albiet finite value as $T \rightarrow 0$~K. This behaviour was found to be robust over a wide range of system parameters~\cite{BaenningerPRL2008,RKoushik_etal_PRB2011}, as well as over a wide range of sample dimensions~\cite{BackesPRB2015,BackesJPCM2016}, only appearing to vanish in the macroscopic system limit~\cite{BaenningerPRL2008}. We stress that this is behaviour is totally at odds with anything observed in macroscopic 2DEGs which, when at comparable $\rho$, are deep within the Anderson localised regime with $\rho$ being strongly $T$-dependent. On the other hand, of the several measured m2DEGs, \textit{not a single one showed any indication of activated behaviour below $\approx 1$~K}. That is, the low-$T$ saturation in $\rho$ was not only seen in a selection of devices, but seen in every device measured.

Figure~\ref{F2}B shows the $T$-dependence of $S$ in the same range over which the saturation in $\rho$ is observed, and it is found that $S$ in m2DEGs grows linearly as a function of $T$ extrapolating to zero at 0~K. This behaviour is qualitatively in agreement with that expected in non-interacting metals (see Eq.~\ref{Mott}). In terms of absolute magnitude, however, $S$ was found to exceed the Mott value by over two orders of magnitude~\cite{NarayanPRB2012}. It is important to note that in the hopping regime $S$ is expected to diverge as $T \rightarrow 0$~K. The precise form of the divergence depends on the precise hopping mechanism, i.e., whether it is nearest-neighbour, variable-ranged or hopping in the presence of the Coulomb gap, but clearly this is totally inconsistent with the experimental observations in Fig.~\ref{F2}b. As $T$ is increased above $\approx 1$~K, $\rho$ appears to decrease in a more conventional activated manner indicating a recovery of the hopping regime. On the other hand $S$ is found to sharply increase as $\sim T^4$ suggesting the role of electron-phonon interactions~\cite{Lyo_PRB1989}. Notably, the rapid~$T^4$-growth sees $S$ attain enormously large values of $\sim 100$~mV/K at 1~K, which are larger than any other previously reported value. 

Thus, to briefly summarise the $T$-dependent characteristics, both $\rho$ and $S$ seem to suggest that the m2DEG has metal-like excitations below $\approx 1$~K	and even at the lowest accessible $\ns$, there seems to be no crossover to strongly localised behaviour. The magnitude of $\rho$ and $S$, however, are significantly larger than is conventionally found in the high-$\ns$ `metallic' regime of 2DEGs and we address this in Section~\ref{Discussion} after inspecting the $\ns$-dependence below.

Figure~\ref{F3} shows the $\ns$-dependence of $\rho$ and $S$~\cite{NarayanPRB2012,NarayanNJP2014}: it is observed that while $\rho(\ns)$ increases essentially in a monotonic fashion, $S$ is markedly non-monotonic showing large oscillations and even sign-changes as a function of $\ns$. The onset of $S$-oscillations was consistently found to occur in the vicinity of $\ns = 1.8 \times 10^{14}~\mbox{m}^{-2}$, above which $S$ was found to agree quantitatively with the Mott prediction (~Eq.~\ref{Mott}). Thus, in contrast with the low-$T$-dependence, the low-$\ns$ behaviour appears to be in qualtitative disagreement with the Mott prediction (Eq.~\ref{MottGeneral}) which requires oscillations in $S$ to be accompanied by oscillations in $\rho$. Notably in this context, $\rho$ was found to be featureless to within the experimental resolution of a few ohms per square~\cite{NarayanNJP2014}. The apparent failure of the Mott picture is further brought out in Fig.~\ref{F4}a where the measured $S$ is compared to $S_{\mbox{Mott}}$ given by Eq.~\ref{MottGeneral}. Not only do $S$ and $S_{\mbox{Mott}}$ oscillate asynchronoulsy, the former exceeds the latter by over two orders of magnitude.

Thus to summarise this section, the experimental data point at an unconventional metallic character in m2DEGs even when $\rho >> h/e^2$, and which is never observed to give way to conventional hopping transport down to the lowest experimentally accessible $\ns$ and/or $T$. That is, the m2DEGs seem to totally avoid the hopping regime below $\approx 1$~K. In addition, $S$ in m2DEGs shows a dramatic departure from the Mott formula. These results are completely novel with no similar findings been reported in macroscopic 2DEGs. In the next section we argue that several of these unexpected results can be understood as arising due to phase coherent electron transport.

\section{Discussion: Manifestations of phase coherence?}
\label{Discussion}

\begin{figure}
	\centering
		\includegraphics[width = 15cm]{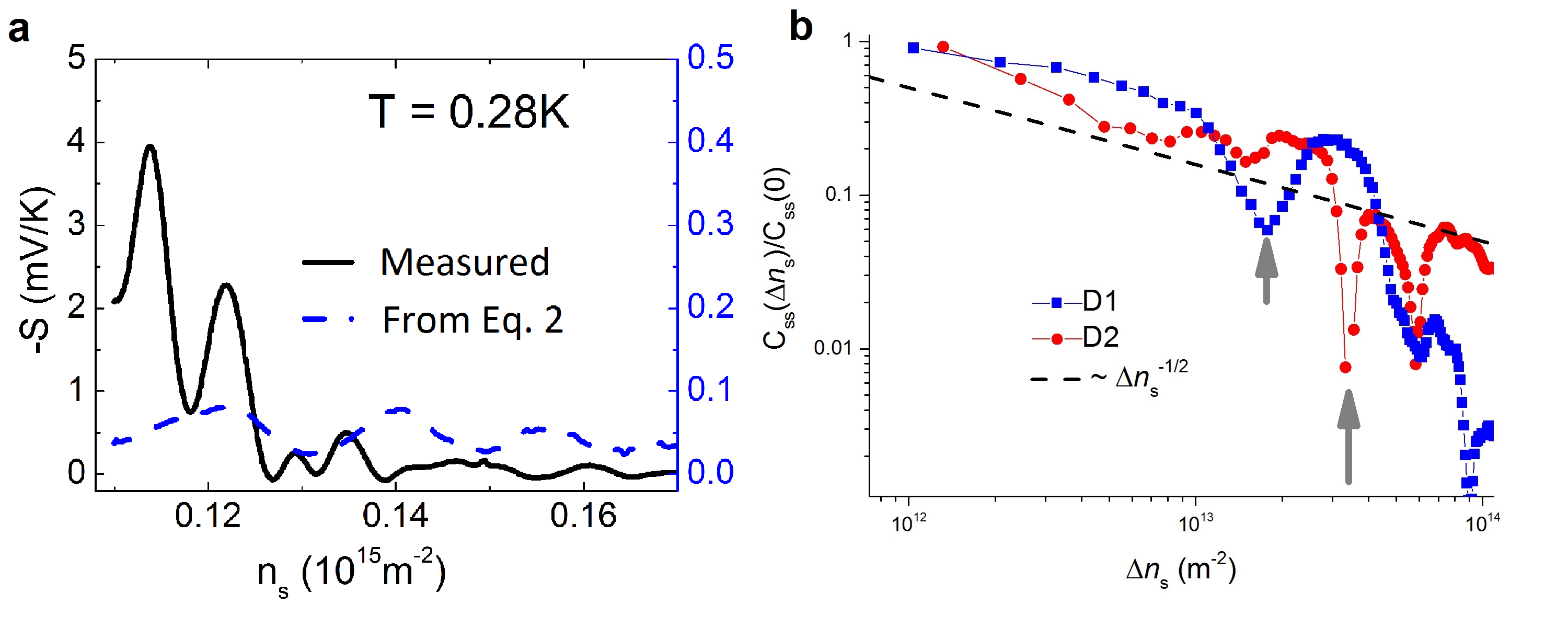}
		\caption{(a) The measured $S$ and that constructed using the measured $\sigma$ and Eq.~\ref{MottGeneral} disagree quantitatively and qualitatively. However, as was shown in Ref.~\cite{LesovikKhmelnitskii} this behaviour is anticipated in phase-coherent systems and is a manifestation of `universal mesoscopic fluctuations'. (b) shows that the two-point correlation function of the $S$ fluctuations (defined as the product $\Delta S(n_{\mbox{s,1}})\Delta S(n_{\mbox{s,2}})$, where $\Delta S(\ns) = S(\ns) - S_{\mbox{d}}(\ns))$ is in strong agreement with the predictions of Ref.~\cite{LesovikKhmelnitskii}, decaying as $\Delta \ns^{-1/2}$, where $\Delta \ns \equiv n_{\mbox{s,1}} - n_{\mbox{s,2}}$. Figure adapted from Ref.~\cite{NarayanNJP2014}.}
	\label{F4}
\end{figure}

One of the most direct indications of phase coherent behaviour is the apperance of `universal mesoscopic fluctuations' which are non-periodic but reproducible fluctuations in the transport as a function of an external parameter such as magnetic field or chemical potential. These arise due to the specific disorder realisation of the system under study which, in mesoscopic systems, tends not be statistically homogeneous. Specifically, the trajectory followed by an electron, and therefore the net phase it accumlates, as it traverses a mesoscopic system, will depend sensitively on scattering events and thus the disorder profile. Since the disorder profile can change in an essentially random manner as the chemical potential is tuned, sharp changes in the electron-interference characteristics can be expected which ultimately manifest as fluctuations in the transport characteristics. In Ref.~\cite{NarayanNJP2014} it was shown that certain aspects of the $S$-oscillations shown in Fig.~\ref{F3}b were strongly consistent with theoretical predictions for mesoscopic $S$ fluctuations~\cite{LesovikKhmelnitskii}. In particular, the autocorrelation of the fluctuations in $S(\ns)$, shown in Fig.~\ref{F4}b, were found to decay as a power law with an exponent of $-1/2$ and showed a minumum before decaying completely, precisely as predicted in Ref.~\cite{LesovikKhmelnitskii}. While a counterpoint to the above explanation could be that no indications of mesoscopic fluctuations are seen in the electrical transport (Fig.~\ref{F3}a), it is noteworthy that even very small (i.e., beyond experimental resolution) but sharp fluctuations in $\sigma$ can engender a dramatic response in $S$. Physically, one can understand this by considering Eq.~\ref{MottGeneral} -- the derivative relation between $S$ and $\sigma$ serves to amplify any fluctuations in the latter and potentially induce divergingly large $S$ in response to discontinous changes in $\sigma$. Importantly the divergences can be negative or positive, thus offering a plausible explanation for the observed sign changes in $S$. 

\begin{figure}
	\centering
		\includegraphics[width = 15cm]{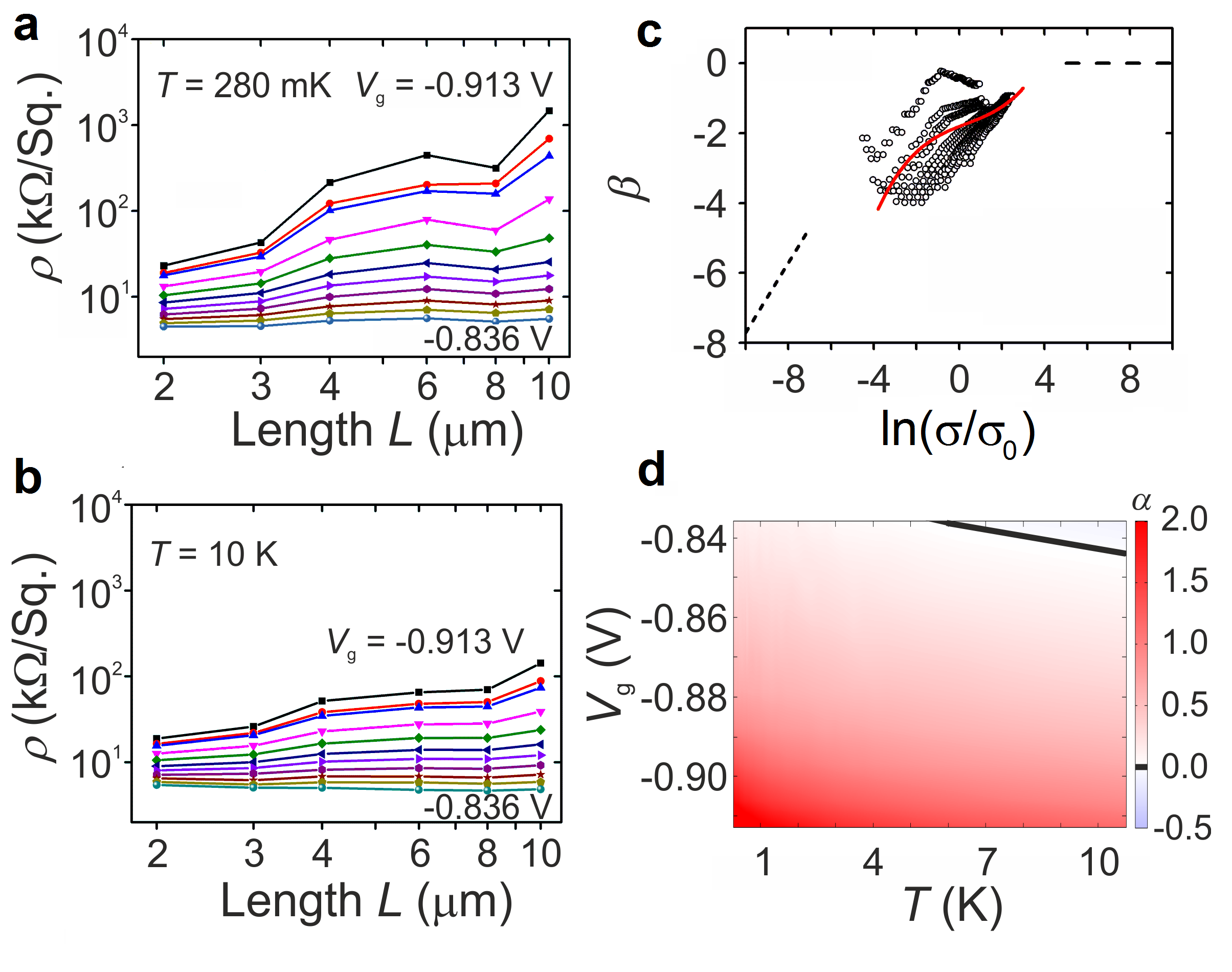}
		\caption{(a) $\rho$ in the m2DEGs shows a very systematic dependence on $L$, increasing roughly in a power-law fashion. (b) This behaviour is seen to persist, albeit to a lesser degree, up to relatively high $T$ of 10~K. From the $L$-dependent characteristics, we construct the `scaling function' $\beta$ (defined in the text) and find it to be in good agreement with the limiting values (shown as broken lines) in the two limits $\sigma >> \sigma_0$ and $\sigma << \sigma_0$, respectively. The red trace is an average of the data points (open circles). (d) Phase diagram of $\alpha$ showing the power-law exponent to increase as $T \rightarrow 0$~K and with increasing $|\Vg|$, i.e., as $\ns$ decreases. Figure adapted from Ref.~\cite{BackesPRB2015}.}
	\label{F5}
\end{figure}

If indeed the mesoscopic size of the 2DEGs causes the fluctuations in $S$, then increasing the 2DEG size should reveal important information. In Ref.~\cite{BackesPRB2015} it was shown that $\rho$ in the m2DEGs had a marked dependence on the system size: as shown in Figs.~\ref{F5}a and~\ref{F5}b the dependence of $\rho$ on $L$ is consistent with a power-law dependence with the power-law exponent $\alpha$ being $T$-dependent. Remarkably, this is precisely the behaviour predicted by the scaling hypothesis of localisation~\cite{AbrahamsetalPRL1979} wherein the finite extent of the electronic wavefunction $\xi$ curbs electron diffusivity on all lengthscales, ultimately manifesting as a geometry-dependent $\sigma$. This is buttressed by Fig.~\ref{F5}c in which we construct from our data the `$\beta$-function' $\equiv \mbox{d}\ln (\sigma/\sigma_0)/\mbox{d}\ln L$, where $\sigma_0 = e^2/h$ and find it to be strongly consistent with the known limiting behaviour when $\sigma << \sigma_0$ and $\sigma >> \sigma_0$. Although system-size-dependent charactertistics have been observed in other systems~\cite{BishopDolanPRL1985,PopovicFowlerWashburnPRL1991}, we believe this is the first direct observation of behaviour consistent with the predictions of the scaling hypothesis. Importantly, the observation of behaviour cosistent with the scaling hypothesis is contingent on a well-defined $\xi$ and therefore on electronic phase coherence over the lengthscales of interest. Thus Fig.~\ref{F5}b is direct evidence of phase coherent transport.

How would this tell on the $T$-dependence of transport? First, as $T$ is increased it is to be expected that phase coherence and, therefore, the $L$-dependent behaviour diminishes. This is clearly seen in Fig.~\ref{F5}d which shows the exponent $\alpha$ to decrease with increasing $T$ (and $|\Vg|$). Perhaps more subtle is the fact that if the electronic wavefunction extends coherently across the m2DEG, then the transport must be metal-like. This can be seen by noting that even though $\xi$ is finite, this is irrelvant in systems with spatial extent $\lesssim \xi$ in which the wavefunction would appear extended. Indeed, this is exactly what is observed in the m2DEGs at $T \lesssim 1$~K with both $\rho$ and $S$ showing behaviour characteristic of metals (Fig.~\ref{F2}a and \ref{F2}b, respectively). But if the system is effectively a metal, why is $\rho$ so large in magnitude? The precise value of $\rho$ depends on the diffusivity of electrons and this is vanishingly small on lengthscales $> \xi$. Importantly, on intermediate lengthscales $\lesssim \xi$ the diffusivity will be diminished, but not zero. It was shown in Ref.~\cite{BackesPRB2015} that the m2DEGs under consideration are precisely in this regime, with $\xi \approx L$. In fact, the low-$T$ saturation value of $\rho$ is observed to increase monotonically with $L$, further supporting the picture of reduced electronic diffusivity with increasing $L$. The last remaining question is then why the metalic behaviour gives way to conventional hopping behaviour above $\approx 1$~K. The point to note here is that irrespective of phase coherence, electrons are continually interacting with phonons, i.e., the hopping transport channel $\sim \exp{(-\Delta/k_B T)}$ is always present. As $T$ is increased, the hopping contribution grows until it exceeds the coherent transport causing a decrese in $\rho(T)$. As shown in Ref.~\cite{BackesPRB2015}, $\rho(T)$ in the m2DEGs can be fitted very well to a simple description in which the hopping and coherent channels constitute two parallel conduction paths.

Thus we see that several of the seemingly anomalous characteristics of the m2DEGs including the metal-like behaviour when $\rho >> h/e^2$ and the apparent breakdown of the Mott relation (Eq.~\ref{MottGeneral}) can be understood as stemming from phase coherent electron transport. However, this is a very surprising conclusion since this would imply $\lphi \sim 10~\mu$m even at 10~K which is many orders of magnitude greater than expected~\cite{Heinzel}. Alternatively, one could ask whether disorder could give rise to the observed effects~\cite{TripathiKennettPRB2006,TripathiKennettPRB2007,NeilsonHamiltonPRB2011}, but the systematic nature of $\rho(L)$ (reported for three independent sets of six m2DEGs each in Ref.~\cite{BackesPRB2015}), the absence of Coulomb blockade features in very narrow m2DEGs~\cite{BackesJPCM2016}, and the consistent observation of the low-$T$ metallic character in numerous 2DEGs~\cite{BaenningerPRL2008,RKoushik_etal_PRB2011,NarayanPRB2012,BackesPRB2015,BackesJPCM2016} strongly argues against disorder-based scenarios. An entirely different argument could be whether the observed phenomena are, as was argued in Refs.~\cite{BaenningerPRL2008,RKoushik_etal_PRB2011,NarayanPRB2012,NarayanJLTP2013}, driven by many-body effects. This is particularly relevant in light of the large magnitude of $S$ since, although intuitively a large $\rho$ also implies a large $S$, the magnitude of $S$ at low $\ns$ ($\lesssim 4 \times 10^{14}~\mbox{m}^{-2}$) is over an order of magnitude larger than predicted by Eq.~\ref{MottGeneral}. In this $\ns$-regime the the interaction parameter $r_s$, defined as the ratio of the Coulomb energy of the systems to its kinetic energy, can be as large as 8 and therefore interactions must play a prominent role. In this respect we point out two theoretical works which predict sign changes~\cite{GargNJP} or divergences~\cite{LimtragoolPRL2014} in $S$ as a function of particle concentration arising due to the proximity to a quantum critical point and strong electron correlations, respectively. However, it is unclear whether these scenarios simultaneously predict a system-size dependence of the transport parameters and/or low-$T$ metal-like transport at very high $\rho$. 

In summary, our recent experimental studies on m2DEGs of spatial extent less than 10~$\mu$m provide strong indications of phase coherence transport even at $T \approx 10$~K. This entirely unexpected result is based on a combination of electrical and thermoelectrical measurements on a wide range of m2DEGs of varying shape and dimension which show (1) metallic $T$-dependence of $\rho$ and $S$ below $\approx 1$~K despite $\rho >> h/e^2$~\cite{BaenningerPRL2008,RKoushik_etal_PRB2011,NarayanPRB2012}, (2) an apparent decoupling of $\rho$ and $S$ wherein the latter oscillates but the former is monotonic as a function of $\ns$~\cite{NarayanJLTP2013,NarayanNJP2014}, and (3) a systematic dependence of $\rho$ on the system-size which is strongly consistent with the predictions of the scaling hypothesis of localisation~\cite{BackesPRB2015}. A consequence of the mesoscopic nature of the 2DEGs is that $S$ can attain enormously large values of up to 100~mV/K at 1.3~K~\cite{NarayanPRB2012} which can be very useful for thermoelectric applications at cryogenic temperatures.




\bibliographystyle{iopart-num} 
\bibliography{references}

\providecommand{\newblock}{}
\begin{thebibliography}{10}
\expandafter\ifx\csname url\endcsname\relax
  \def\url#1{{\tt #1}}\fi
\expandafter\ifx\csname urlprefix\endcsname\relax\def\urlprefix{URL }\fi
\providecommand{\eprint}[2][]{\url{#2}}

\bibitem{Wharam_etal_JPhysC1988}
Wharam D~A, Thornton T, Newbury R, Pepper M, Ahmed H, Frost J~E~F, Hasko D~G,
  Peacockt D~C, Ritchie D~A and Jones G~A~C 1988 {\em J. Phys. C: Sol. Stat.
  Phys.\/} {\bf 21} L209

\bibitem{vanWees_etal_PRL1988}
van Wees B~J, van Houten H, Beenakker C~W~J, Williamson J~G, Kouwenhoven L~P,
  van~der Marel D and Foxon C~T 1988 {\em PRL\/} {\bf 60} 848

\bibitem{Smith_etal_JPhysC1988}
Smith C, Pepper M, Ahmed H, Frost J, Hasko D, Peacock D, Ritchie D and Jones G
  1988 {\em J. Phys. C: Sol. Stat. Phys.\/} {\bf 21} L893

\bibitem{vanWees_etal_PRL1989}
van Wees B, Kouwenhoven L, Harmans C, Williamson J, Timmering C, Broekaart M,
  Foxon C and Harris J 1989 {\em PRL\/} {\bf 62} 2523

\bibitem{Simmons_etal_PRL2000}
Simmons M~Y, Hamilton A~R, M~Pepper~and E~H~L, Rose P~D and Ritchie D~A 2000
  {\em Phys. Rev. Lett.\/} {\bf 84} 2489

\bibitem{Hamilton_etal_PRL2000}
Hamilton A~R, Simmons M~Y, Pepper M, Linfield E~H and Ritchie D~A 2000 {\em
  PRL\/} {\bf 87} 126802

\bibitem{Abrahams_etal_RMP2001}
Abrahams E, Kravchenko S~V and Sarachik M~P 2001 {\em Rev. Mod. Phys.\/} {\bf
  73} 251

\bibitem{IQHE}
v~Klitzing K, Dorda G and Pepper M 1980 {\em PRL\/} {\bf 45} 494

\bibitem{FQHE}
Tsui D~C, Stormer H~L and Gossard A~C 1982 {\em PRL\/} {\bf 48} 1559

\bibitem{BackesPRB2015}
Backes D, Hall R, Pepper M, Beere H, Ritchie D and Narayan V 2015 {\em PRB\/}
  {\bf 92} 235427

\bibitem{NarayanNJP2014}
Narayan V, Kogan E, Ford C, Pepper M, Kaveh M, Griffiths J, Jones G, Beere H
  and Ritchie D 2014 {\em New J. Phys.\/} {\bf 16} 085009

\bibitem{NarayanPRB2012}
Narayan V, Pepper M, Griffiths J, Beere H, Sfigakis F, Jones G, Ritchie D and
  Ghosh A 2012 {\em Phys. Rev. B\/} {\bf 86} 125406

\bibitem{Heinzel}
Heinzel T 2007 {\em Mesoscopic Electronics in Solid State Nanostructures\/}
  (Weinheim: Wiley-VCH)

\bibitem{BillialdAPL2015}
Billiald J, Backes D, K{\"o}nig J, Farrer I, Ritchie D and Narayan V 2015 {\em
  Appl. Phys. Lett.\/} {\bf 107} 022104

\bibitem{Syme_etal_JPhys1989}
Syme R, Kelly M and Pepper M 1989 {\em J. Phys.: Condens. Matter\/} {\bf 1}
  3375

\bibitem{ChickeringPRL2009}
Chickering W~E, Eisenstein J~P and Reno J~L 2009 {\em Phys. Rev. Lett.\/} {\bf
  103} 046807

\bibitem{MottDavis}
Mott N~F and Davis E~A 1971 {\em Electronic Processes in Non-crystalline
  Materials\/} (Oxford: Clarendon Press)

\bibitem{NarayanJLTP2013}
Narayan V, Pepper M, Griffiths J, Beere H, Sfigakis F, Jones G, Ritchie D and
  Ghosh A 2013 {\em J. Low Temp. Phys.\/} {\bf 171} 626

\bibitem{Fletcher_etal_SST2001}
Fletcher R, Pudalov V~M, Radcliffe A~D~B and Possanzini C 2001 {\em Semicond.
  Sci. Technol.\/} {\bf 16} 386

\bibitem{AbrahamsetalPRL1979}
Abrahams E, Anderson P~W, Licciardello D~C,  and Ramakrishnan T~V 1979 {\em
  PRL\/} {\bf 42} 673

\bibitem{LeeRamakrishnanRMP1985}
Lee P~A and Ramakrishnan T~V 1985 {\em Rev. Mod. Phys.\/} {\bf 57} 287

\bibitem{SDSarma_etal_PRL2005}
Sarma S~D, Lilly M~P, Hwang E~H, Pfeiffer L~N, West K~W and Reno J~L 2005 {\em
  PRL\/} {\bf 94} 136401

\bibitem{BaenningerPRL2008}
Baenninger M, Ghosh A, Pepper M, Beere H~E, Farrer I and Ritchie D~A 2008 {\em
  Phys. Rev. Lett.\/} {\bf 100} 016805

\bibitem{RKoushik_etal_PRB2011}
Koushik R, Baenninger M, Narayan V, Mukerjee S, Pepper M, Farrer I, Ritchie D~A
  and Ghosh A 2011 {\em Phys. Rev. B\/} {\bf 83} 085302

\bibitem{BackesJPCM2016}
Backes D, Hall R, Pepper M, Beere H, Ritchie D and Narayan V 2016 {\em J.
  Phys.: Cond. Mat.\/} {\bf 28} 1LT01

\bibitem{Lyo_PRB1989}
Lyo S~K 1989 {\em Phys. Rev. B (R)\/} {\bf 38} 6345

\bibitem{LesovikKhmelnitskii}
GB~Lesovik D~K 1988 {\em Sov. Phys. JETP.\/} {\bf 67} 957

\bibitem{BishopDolanPRL1985}
Bishop D~J and Dolan G~J 1985 {\em PRL\/} {\bf 55} 2911

\bibitem{PopovicFowlerWashburnPRL1991}
Popovi{\'c} D, Fowler A and Washburn S 1992 {\em PRL\/} {\bf 67} 2870

\bibitem{TripathiKennettPRB2006}
Tripathi V and Kennett M~P 2006 {\em Phys. Rev. B\/} {\bf 74} 195334

\bibitem{TripathiKennettPRB2007}
Tripathi V and Kennett M~P 2007 {\em Phys. Rev. B\/} {\bf 76} 115321

\bibitem{NeilsonHamiltonPRB2011}
Neilson D and Hamilton A~R 2011 {\em Phys. Rev. B\/} {\bf 84} 129901

\bibitem{GargNJP}
Garg A, Shastry B~S, Dave K~B and Phillips P 2011 {\em New J. Phys.\/} {\bf 13}
  083032

\bibitem{LimtragoolPRL2014}
Limtragool K and Phillips P~W 2014 {\em PRL\/} {\bf 113} 086405

\end{thebibliography}





\end{document}